\documentstyle[preprint,tighten,aps,floats,psfig]{revtex}
\def\ppbar{$p\overline{p} $}            
\def\ttbar{$t\overline{t}$}             
\def\met{\mbox{${\hbox{$E$\kern-0.6em\lower-.1ex\hbox{/}}}_T$}} 
\def\ipb{pb$^{-1}$}                     
\def\gevcc{GeV/$c^2$}                   
\def\D0{D\O}                            
\def\etal{{\sl et al.}}                 
\begin{document}
\preprint{\large Fermilab-Pub-97/344-E}
\draft
\title{\Large\bf
Search for First Generation Scalar Leptoquark Pairs in \ppbar\
Collisions at $\sqrt s$ = 1.8 TeV}
\date{October 29, 1997}
%
%
\author{                                                                      
B.~Abbott,$^{30}$                                                             
M.~Abolins,$^{27}$                                                            
B.S.~Acharya,$^{45}$                                                          
I.~Adam,$^{12}$                                                               
D.L.~Adams,$^{39}$                                                            
M.~Adams,$^{17}$                                                              
S.~Ahn,$^{14}$                                                                
H.~Aihara,$^{23}$                                                             
G.A.~Alves,$^{10}$                                                            
E.~Amidi,$^{31}$
% ...
\\                                                                            
\vskip 0.50cm                                                                 
\centerline{(D\O\ Collaboration)}                                             
\vskip 0.50cm                                                                 
}                                                                             
\address{                                                                     
\centerline{$^{1}$Universidad de los Andes, Bogot\'{a}, Colombia}             
\centerline{$^{2}$University of Arizona, Tucson, Arizona 85721}               
\centerline{$^{3}$Boston University, Boston, Massachusetts 02215}             
\centerline{$^{4}$Brookhaven National Laboratory, Upton, New York 11973}      
\centerline{$^{5}$Brown University, Providence, Rhode Island 02912}           
\centerline{$^{6}$Universidad de Buenos Aires, Buenos Aires, Argentina}       
\centerline{$^{7}$University of California, Davis, California 95616}          
\centerline{$^{8}$University of California, Irvine, California 92697}         
\centerline{$^{9}$University of California, Riverside, California 92521}      
\centerline{$^{10}$LAFEX, Centro Brasileiro de Pesquisas F{\'\i}sicas,        
                  Rio de Janeiro, Brazil}                                     
\centerline{$^{11}$CINVESTAV, Mexico City, Mexico}                            
\centerline{$^{12}$Columbia University, New York, New York 10027}             
\centerline{$^{13}$Delhi University, Delhi, India 110007}                     
\centerline{$^{14}$Fermi National Accelerator Laboratory, Batavia,            
                   Illinois 60510}                                            
\centerline{$^{15}$Florida State University, Tallahassee, Florida 32306}      
\centerline{$^{16}$University of Hawaii, Honolulu, Hawaii 96822}              
\centerline{$^{17}$University of Illinois at Chicago, Chicago,                
                   Illinois 60607}                                            
\centerline{$^{18}$Indiana University, Bloomington, Indiana 47405}            
\centerline{$^{19}$Iowa State University, Ames, Iowa 50011}                   
\centerline{$^{20}$Korea University, Seoul, Korea}                            
\centerline{$^{21}$Kyungsung University, Pusan, Korea}                        
\centerline{$^{22}$Institute of Nuclear Physics, Krak\'ow, Poland}            
\centerline{$^{23}$Lawrence Berkeley National Laboratory and University of    
                   California, Berkeley, California 94720}                    
\centerline{$^{24}$Louisiana Tech University, Ruston, Louisiana 71272}        
\centerline{$^{25}$University of Maryland, College Park, Maryland 20742}      
\centerline{$^{26}$University of Michigan, Ann Arbor, Michigan 48109}         
\centerline{$^{27}$Michigan State University, East Lansing, Michigan 48824}   
\centerline{$^{28}$Moscow State University, Moscow, Russia}                   
\centerline{$^{29}$University of Nebraska, Lincoln, Nebraska 68588}           
\centerline{$^{30}$New York University, New York, New York 10003}             
\centerline{$^{31}$Northeastern University, Boston, Massachusetts 02115}      
\centerline{$^{32}$Northern Illinois University, DeKalb, Illinois 60115}      
\centerline{$^{33}$Northwestern University, Evanston, Illinois 60208}         
\centerline{$^{34}$University of Notre Dame, Notre Dame, Indiana 46556}       
\centerline{$^{35}$University of Oklahoma, Norman, Oklahoma 73019}            
\centerline{$^{36}$University of Panjab, Chandigarh 16-00-14, India}          
\centerline{$^{37}$Institute for High Energy Physics, 142-284 Protvino,       
                   Russia}                                                    
\centerline{$^{38}$Purdue University, West Lafayette, Indiana 47907}          
\centerline{$^{39}$Rice University, Houston, Texas 77005}                     
\centerline{$^{40}$Universidade do Estado do Rio de Janeiro, Brazil}          
\centerline{$^{41}$University of Rochester, Rochester, New York 14627}        
\centerline{$^{42}$CEA, DAPNIA/Service de Physique des Particules,            
                   CE-SACLAY, Gif-sur-Yvette, France}                         
\centerline{$^{43}$Seoul National University, Seoul, Korea}                   
\centerline{$^{44}$State University of New York, Stony Brook,                 
                   New York 11794}                                            
\centerline{$^{45}$Tata Institute of Fundamental Research,                    
                   Colaba, Mumbai 400005, India}                              
\centerline{$^{46}$University of Texas, Arlington, Texas 76019}               
\centerline{$^{47}$Texas A\&M University, College Station, Texas 77843}       
\vspace{0.2in}
}                                                                             
%
\maketitle
\newpage
\vspace*{1.5in}
\begin{abstract}
We have searched for  first generation scalar  leptoquark (LQ) pairs in the
$e\nu$+jets    channel  using  \ppbar\   collider  data  ($\int  Ldt\approx
115$~pb$^{-1}$)  collected by the D\O\ experiment  at the Fermilab Tevatron
during 1992--96.  The analysis  yields no candidate  events. We combine the
results with those from the  $ee$+jets and $\nu\nu$+jets channels to obtain
95\%  confidence level (CL)  upper limits on  the LQ pair  production cross
section as a function of  mass and of $\beta$,  the branching fraction to a
charged  lepton. Comparing  with the  next-to-leading order  theory, we set
95\% CL  lower  limits  on the  LQ mass  of 225,  204,  and  79~\gevcc\ for
$\beta=1$, ${1 \over 2}$, and 0, respectively.
\end{abstract}
\vskip 0.2in
\pacs{PACS numbers: 12.60.-i, 12.90.+b, 14.80.-j, 13.85.Rm\\[0.3in]
{\it Submitted to Physical Review Letters}
}
%
\newpage

One of the remarkable  features of the Standard  Model (SM) is the symmetry
between quarks and  leptons that leads to  cancellation of chiral anomalies
and renders the  SM renormalizable.  This symmetry  might be explained by a
more    fundamental  theory  that  relates   quarks and   leptons.  Several
extensions~\cite{generic_lq}      of  the  SM  include    leptoquarks (LQ),
color-triplet   bosons which  carry both   lepton~($\ell$) and  quark~($q$)
quantum  numbers. The  masses and  coupling  strengths of  leptoquarks that
couple to all  three fermion  generations  are severely  constrained by low
energy experiments~\cite{low_e} and by HERA~\cite{HERA1}. Therefore, only LQ
that couple to a single generation  can be light enough to be accessible at
present    accelerators. The  excess  of  events at  high  $Q^2$  in $e^+p$
collisions reported~\cite{hera} by the H1 and ZEUS experiments at HERA, and
its possible   interpretation~\cite{hewett}  as evidence  for production of
first  generation scalar  leptoquarks with a  mass near  $200$~\gevcc, have
stimulated much interest in these particles. 

Leptoquarks would  be dominantly  pair-produced via  strong interactions in
\ppbar\  collisions,  independently of the  unknown  LQ--$\ell$--$q$ Yukawa
coupling.  Each  leptoquark would  subsequently  decay into a  lepton and a
quark. For first generation leptoquarks, this leads to three possible final
states: $ee$+jets,  $e\nu$+jets and  $\nu\nu$+jets, with rates proportional
to $\beta^2$,   $2\beta(1-\beta)$, and  $(1-\beta)^2$,  respectively, where
$\beta$ denotes the branching fraction of a leptoquark to an electron and a
quark (jet). The  CDF~\cite{old_lq1,cdflq} and  D\O\ \cite{old_lq2,eejjprl}
Collaborations have both searched  for first generation scalar leptoquarks.
The recent    analyses\cite{eejjprl,cdflq} of the  $ee$+jets  decay channel
yielded 95\% confidence  level (CL) lower limits  on the leptoquark mass of
225~\gevcc\  (D\O)  and  213~\gevcc\  (CDF). In this  Letter we  present an
analysis of the  $e\nu$+jets final state (which  has maximum sensitivity at
$\beta=  {1 \over  2}$),  using $115  \pm   6$~pb$^{-1}$\ of  collider data
collected at the Fermilab Tevatron at $\sqrt{s} = 1.8$ TeV during 1992--96.
We also present a reinterpretation of our search~\cite{stop} for top squark
pairs as a search  for leptoquarks  in the  $\nu\nu$+jets decay channel. We
combine the results from all three decay channels to obtain lower limits on
the leptoquark mass as a function of $\beta$.

The  D\O\  detector  \cite{d0nim}  consists  of a  central  tracking system
including  a   transition   radiation   detector, a    uranium/liquid-argon
calorimeter and a  muon  spectrometer. The data used in  this analysis were
collected with triggers  which required the  presence of an electromagnetic
object,  with or without  jets and  missing  transverse  energy (\met). The
combined  efficiency of  the  triggers is greater  than 98\%  for LQ masses
above  80~\gevcc.  Off\-line event  selection  requires: one  electron with
transverse   energy   $E_T^e>20$~GeV and   pseudorapidity   $|\eta|<1.1$ or
$1.5<|\eta|<2.5$; $\met>30$~GeV as the signature for a neutrino; and two or
more jets  reconstructed  using a  cone algorithm  (cone  radius ${\cal{R}}
\equiv  \sqrt{(\Delta\phi)^2  +(\Delta\eta)^2}= 0.7 $,  where $\phi$ is the
azimuthal angle) with $E_T^j >  20$~GeV and $|{\eta}| < 2.5$. To reduce the
effects of jet  energy  mismeasurements, we require the  \met\ vector to be
separated from the jets by $\Delta\phi > 0.25$ radians if $\met < 120$~GeV.
To  suppress  background from  heavy  quark events,  we reject  events that
contain muons. These selection criteria define our base sample.

An  electron is   identified by its  pattern  of energy   deposition in the
calorimeter,  the  presence of  a  matching track in  the  central tracking
detectors,  and  ionization in the  central  detectors. The  efficiency for
finding   an   electron  is   calculated  to be     (61$\pm$4)\%,  using $Z
$($\rightarrow   ee$)+jets  events  which are  similar  in  topology to the
signal. All  kinematic  quantities  in the event  are  calculated using the
event vertex determined by the electron. 

We use the   {\footnotesize{ISAJET}}  \cite{isajet} Monte  Carlo (MC) event
generator to simulate LQ signal events for masses ($M_{\rm LQ}$) between 80
and   220~\gevcc\  in   20~\gevcc\  steps.  The     {\footnotesize{PYTHIA}}
\cite{pythia}  MC program is  used to  simulate signal  events with $M_{\rm
LQ}$= 200~\gevcc\  to study  systematic errors arising  from differences in
modeling of the  gluon radiation and  parton  fragmentation. The leptoquark
production cross sections used are  from recent next-to-leading order (NLO)
calculations~\cite{kraemer}. The  dominant $W$+jets background is simulated
using the  {\footnotesize{VECBOS}} \cite{vecbos}  event generator (with the
{\footnotesize{HERWIG}}~\cite{herwig}    program  used for  fragmenting the
partons). The  background  from multijet  events containing  a jet which is
misidentified    as  an    electron,  and  with  \met\   arising   from the
mismeasurement   of jet   energies, is  modeled  using  multijet  data. The
probability   for a  jet to  be   misidentified as  an  electron  (the fake
probability)  is    estimated\cite{eejjprl}  to be  $(3.50 \pm  0.35)\times
10^{-4}$. The background from \ttbar\ decays into one or two electrons plus
two or more jets, is simulated using the {\footnotesize{HERWIG}} MC program
with a top  quark mass of  170~\gevcc. All  MC event  samples are processed
through       the   D\O\       detector       simulation    based   on  the
{\footnotesize{GEANT}}~\cite{geant} package. 

In the base sample of 1094 events, we estimate the number of \ttbar\ events
to be $12 \pm  4$ using the  measured  \ttbar\ production  cross section of
$5.5 \pm 1.8$~pb~\cite{top_sigma}.  The multijet background is estimated to
be $75 \pm 15$  events, using  a sample of  events with  three or more jets
with $\met  >30$~GeV. This is done  by multiplying the  fake probability by
the number of  ways the events  satisfy the selection  criteria with one of
the jets  passing the  electron  $E_T$ and $\eta$   requirements. After the
estimated numbers of \ttbar\ and multijet background events are subtracted,
the number  of events  with  transverse mass of  the electron  and neutrino
($M_T^{e\nu}$)     below    110~\gevcc\  is  used to   obtain  an  absolute
normalization for the $W$+jets  background. This background is then largely
eliminated by requiring  $M_T^{e\nu}>110$~\gevcc. After this cut, 14 events
remain in the final data sample. The estimated background is $17.8 \pm 2.1$
events, of which  $11.7 \pm 1.8$,  $4.1 \pm 0.9$, and  $2.0 \pm 0.7$ events
are  from  $W$+jets,  multijets,  and  \ttbar\   production,  respectively.
Leptoquark pair production would yield 24 events in this sample, if $M_{\rm
LQ}=120$~\gevcc\  and $\beta  ={1 \over 2}$.  Assuming all  14 events to be
signal, LQ production  for masses below 120  \gevcc\ can be excluded at the
95\% CL for $\beta ={1 \over 2}$ with no further optimization.

We have  identified  two  additional  variables that  provide   significant
discrimination  between signal and  the remaining  background. They are the
scalar transverse  energy sum $S_T  \equiv E_T^e +  E_T^{j_1} + E_T^{j_2} +
\met$, where $E_T^{j_{1,2}}$ are the transverse energies of the two leading
jets,  and  a  mass   variable     $\frac{dM}{M}(M_{\rm  LQ})   \equiv {\rm
min}(\frac{|M_{ej_1}-M_{\rm   LQ}|}  {M_{\rm  LQ}},  \frac{|M_{ej_2}-M_{\rm
LQ}|}{M_{\rm    LQ}})$,  where  $M_{\rm  LQ}$  is an   assumed LQ  mass and
$M_{ej_{1,2}}$ are the invariant  masses of the electron with the first and
second leading jets.

\begin{figure}[p]
\vbox{
\centerline{\psfig{figure=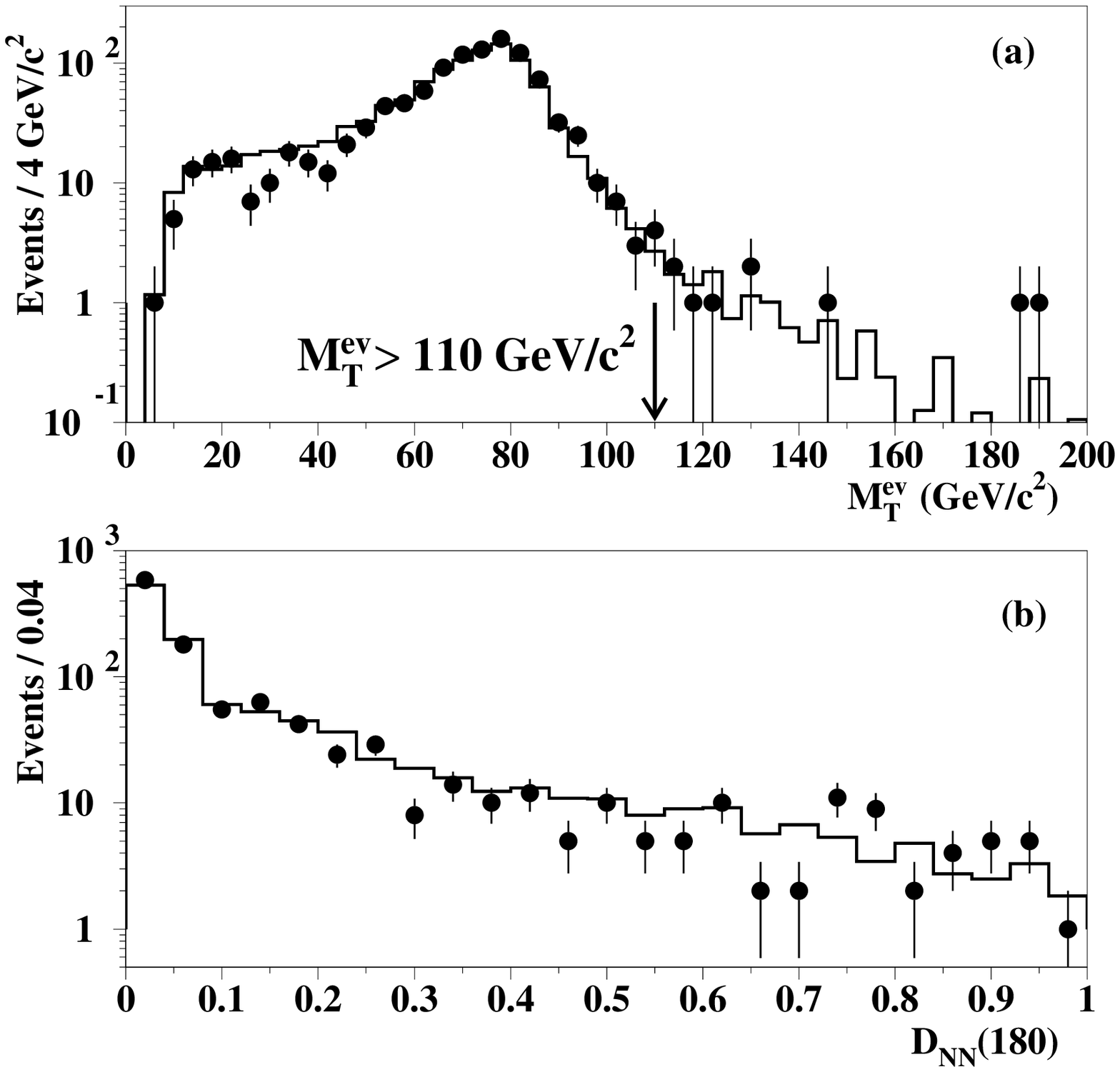,width=6.5in}}
\caption{(a)  $M_T^{e\nu}$ and (b)  ${\cal D}_{\rm  NN}(180)$ distributions
for data  (points) and  background  (histograms), after all  cuts except on
$M_T^{e\nu}$ and  ${\cal D}_{\rm  NN}(180)$. The arrow  in plot (a) shows a
cut on $M_T^{e\nu}$, as described in the text.} \label{fig:Bck_comp}}
\end{figure}

To find the optimal  selection cuts, we adopt  the criterion~\cite{eejjprl}
of maximizing the MC signal  efficiency for a fixed expected background  of
approximately     0.4~events.  In  the low  mass  range   ($M_{\rm  LQ} \le
120$~\gevcc ), where LQ production rates are high, requiring $S_T >400$~GeV
is sufficient. For $M_{\rm LQ} >  120$ \gevcc , we use neural networks (NN)
since they provide higher efficiency than an $S_T$ cut alone. At each mass,
$M_{\rm  LQ}$,  where we  have  generated MC  events, we use  a three layer
feed-forward    neural   network   \cite{nn} with  two   inputs  ($S_T$ and
$\frac{dM}{M}(M_{\rm   LQ})$), five  hidden  nodes, and one  output (${\cal
D}_{\rm NN}(M_{\rm  LQ}$)).  We train each NN  using simulated LQ events as
the signal (with desired ${\cal D}_{\rm NN}(M_{\rm LQ}) = 1$) and a mixture
of  $W$+jets,  multijet, and  \ttbar\  events as  background  (with desired
${\cal D}_{\rm  NN}(M_{\rm  LQ})$~=~0). Cuts  on ${\cal  D}_{\rm NN}(M_{\rm
LQ})$ that yield background estimates closest to the desired background are
obtained by varying ${\cal D}_{\rm  NN}(M_{\rm LQ})$ in steps of 0.05.  The
background  after the  cut ranges  between $0.29  \pm 0.25$   and $0.61 \pm
0.27$  events  as  shown in     Table~\ref{tab:table1}. The   errors in the
background  estimates include  the effects  of  uncertainties in jet energy
scale, fake  probability and \ttbar\  production cross  section. The signal
detection  efficiencies  calculated using  simulated LQ  events passing the
selection requirements are also shown in Table~\ref{tab:table1}. The errors
on the signal  efficiencies include  uncertainties in  trigger and particle
identification  efficiencies, jet energy scale,  effects of gluon radiation
and parton fragmentation in the  signal modeling, and finite MC statistics.
No data events pass the cuts.

To demonstrate  that the backgrounds  are reliably  modeled, comparisons of
the data and combined  background in the  variables $M_T^{e\nu}$ and ${\cal
D}_{\rm NN}(180)$ are shown in Fig.~\ref{fig:Bck_comp} for the base sample.

\begin{figure}[p]
\vbox{
\centerline{\psfig{figure=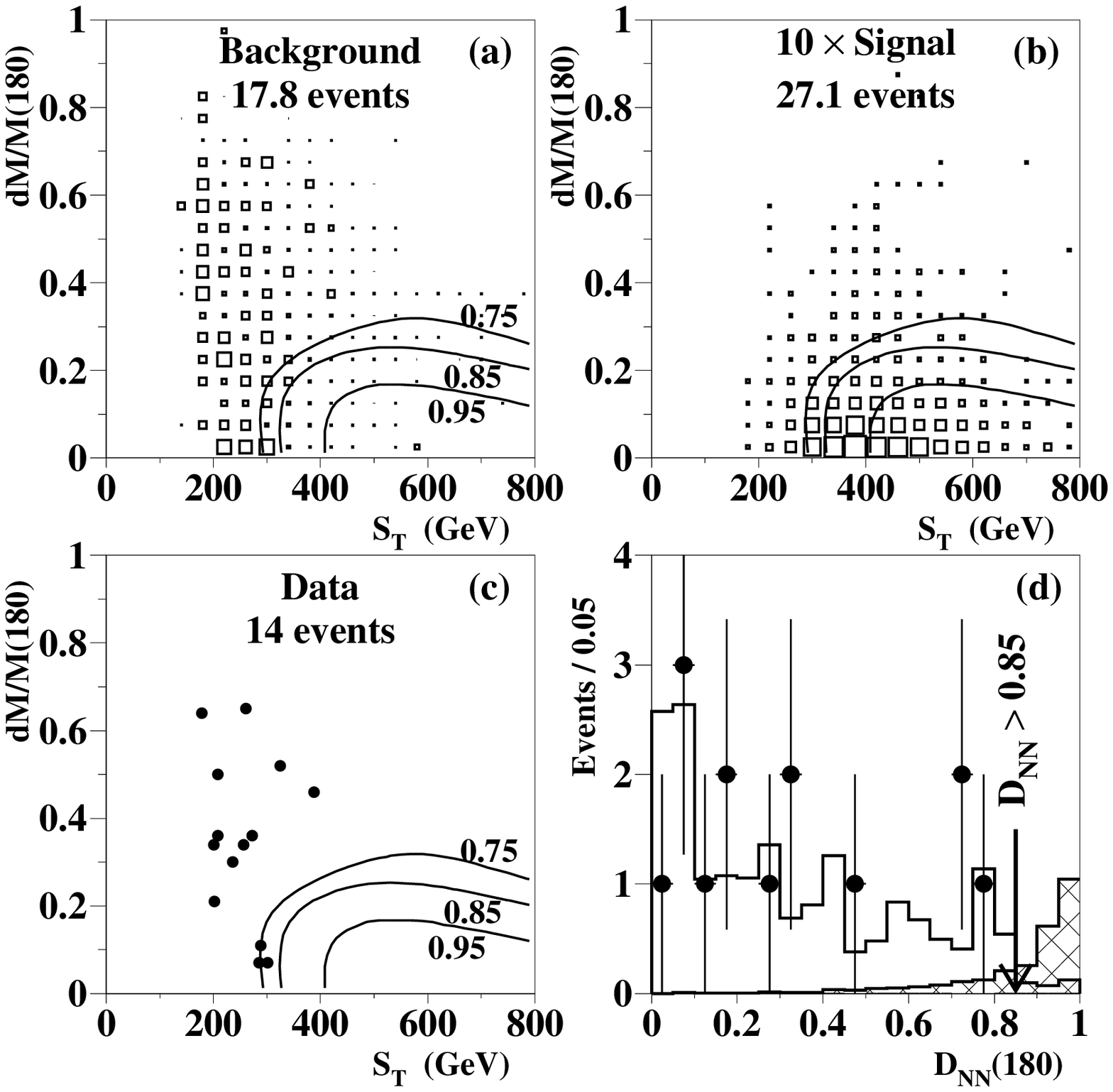,width=6.5in}}
\caption{  Distributions of   $\frac{dM}{M}(180)$ {\sl  vs.}  $S_T$  for (a)
predicted background, (b) simulated  LQ events ($M_{\rm LQ}$ = 180 \gevcc),
and (c) data, after all  cuts except that on  ${\cal D}_{\rm NN}(180)$. The
contours correspond to ${\cal D}_{\rm NN}(180) = 0.75$, 0.85, and 0.95. The
box area is proportional to the  number of events in the bin, with the
total number of events normalized to 115 \ipb. Plot (d) shows distributions
of ${\cal  D}_{\rm  NN}(180)$ for  data (solid  circles),  background (open
histogram) and expected  LQ signal for $M_{\rm  LQ}$ = 180 \gevcc\ (hatched
histogram). The arrow in plot (d) shows the chosen cut on ${\cal  D}_{\rm
NN}(180)$, as described in the text.}\label{fig:nnstdm}}
\end{figure}

Figures~\ref{fig:nnstdm}~(a)--(c)  show the  2-dimensional distributions of
$\frac{dM}{M}(180)$  {\sl  vs.} $S_T$ for  simulated LQ  signal events with
$M_{\rm LQ} = 180$~\gevcc, the  combined background, and data. The contours
corresponding to  constant values of  ${\cal D}_{\rm  NN}(180)$ demonstrate
the separation achieved between  signal and background. The distribution of
${\cal   D}_{\rm   NN}(180)$  for  data is   compared  with  the  predicted
distributions for background and signal in Fig.~\ref{fig:nnstdm}~(d). It is
clear that the  data are  described well by  background  alone. The highest
${\cal D}_{\rm NN}(180)$ observed in the final data sample is 0.79. 

\begin{figure}[p]
\vbox{
\centerline{
\psfig{figure=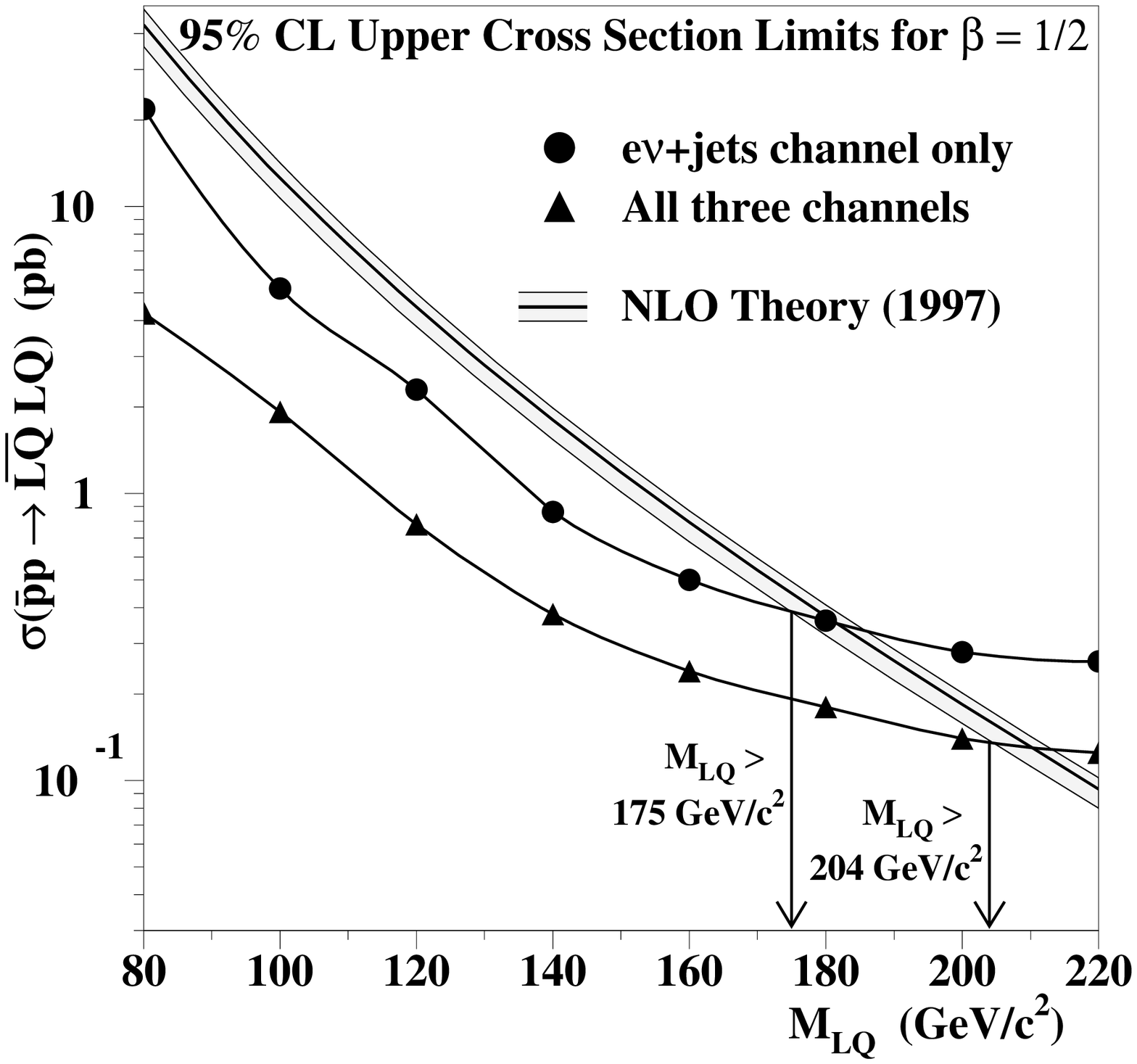,width=6.5in}}
\caption{ Measured 95\%  CL upper limits on the  leptoquark pair production
cross section (see text) 
in  the $e\nu + {\rm  jets}$ channel  (circles) and all three
channels combined (triangles) for  $\beta= {1 \over 2}$. Also shown are the
NLO  calculations of    Ref.~\protect\cite{kraemer} where the  central line
corresponds to $\mu = M_{\text{\rm  LQ}}$, and the lower and upper lines to
$\mu = 2M_{\text{\rm LQ}}$ and $\mu = \frac{1}{2}M_{\rm LQ}$, respectively. }
\label{fig:limit} }
\end{figure}

Using   Bayesian   statistics,  we  obtain a  95\% CL  upper  limit  on the
leptoquark  pair  production cross  section for  $\beta= {1  \over 2}$ as a
function     of      leptoquark    mass.   The    results   are    shown in
Table~\ref{tab:table1}. The statistical and systematic uncertainties in the
efficiency, the  integrated  luminosity, and the  background estimation are
included in the  limit calculation  with Gaussian prior  probabilities. The
measured  95\%  CL cross   section upper  limits  for LQ  pair  production,
corrected for the  branching ratio with $\beta=  {1 \over 2}$,  for various
LQ  masses are  plotted  in    Fig.~\ref{fig:limit}  together  with the NLO
calculations~\cite{kraemer}.  The intersection of  the limit curve with the
lower edge of the theory  band (renormalization  scale $\mu = 2M_{\text{\rm
LQ}}$) is at $ 0.19$ pb, leading to a 95\% CL lower limit on the LQ mass of
175 \gevcc.

\begin{table}[p]
\caption{Signal detection  efficiencies, estimated backgrounds and measured
95\% CL upper limits on the  production cross section from the $e\nu$+jets
channel  analysis. The NLO  cross sections (with  $\mu=2M_{\text{\rm
LQ}}$) from Ref.~\protect\cite{kraemer} times $2\beta(1-\beta)=\frac{1}{2}$
for $\beta = {1 \over 2}$ are also shown. }
\vskip 0.1in
\begin{tabular} { c c c c c }
Leptoquark  & Signal  & Estimated &
95\% CL  & NLO   \\
Mass & Efficiency & Background & Upper Limit & Theory \\
(\gevcc) & (\%) & (Events) & (pb) & (pb) \\
\hline
80 & $0.3 \pm 0.1$  & $0.60 \pm 0.27$ & 10.88 & 17.98 \\
100 & $1.2 \pm 0.2$ & $0.60 \pm 0.27$ & 2.59 & 5.34 \\
120 & $2.5 \pm 0.3$  & $0.60 \pm 0.27$ & 1.15 & 1.90 \\
140 & $6.7 \pm 1.0$  & $0.54 \pm 0.25$ & 0.43 & 0.77 \\
160 & $10.9 \pm 1.2$ & $0.61 \pm 0.27$ & 0.25 & 0.34 \\
180 & $14.7 \pm 1.2$ & $0.29 \pm 0.25$ & 0.18 & 0.16 \\
200 & $19.4 \pm 1.7$ & $0.43 \pm 0.27$ & 0.14 & 0.08 \\
220 & $21.5 \pm 1.7$ & $0.41 \pm 0.27$ & 0.13 & 0.04 \\
\end{tabular}
\label{tab:table1}
\end{table}

An analysis of the $\nu\nu + {\rm  jets}$ channel is accomplished by making
use of our published search (with  $\int Ldt\approx 7.4$~pb$^{-1}$) for the
supersymmetric partner  of the top quark  \cite{stop}. Three events survive
the selection criteria ($\met  >40$~GeV, 2 jets with $E_T^j>30$~GeV, and no
isolated  electrons or muons)  consistent with the  estimated background of
$3.5\pm 1.2$ events, mainly from  $W$/$Z$+jets production. The efficiencies
of the  event  selection  for  $M_{\rm LQ}$=  60, 80,  and  100~\gevcc\ are
calculated to  be 1.1\%,  2.2\%, and 3.9\%,  respectively,  using signal MC
events generated with  the  {\footnotesize{ISAJET}} generator and processed
through  the  detector  simulation  based on    {\footnotesize{GEANT}}. The
systematic    errors  in  the  signal   acceptance  are   calculated  as in
Ref.~\cite{stop}. This analysis yields the limit $M_{\rm LQ}>$79~\gevcc\ at
the 95\% CL for $\beta=$0.

\begin{figure}[p]
\vbox{
\centerline{
\psfig{figure=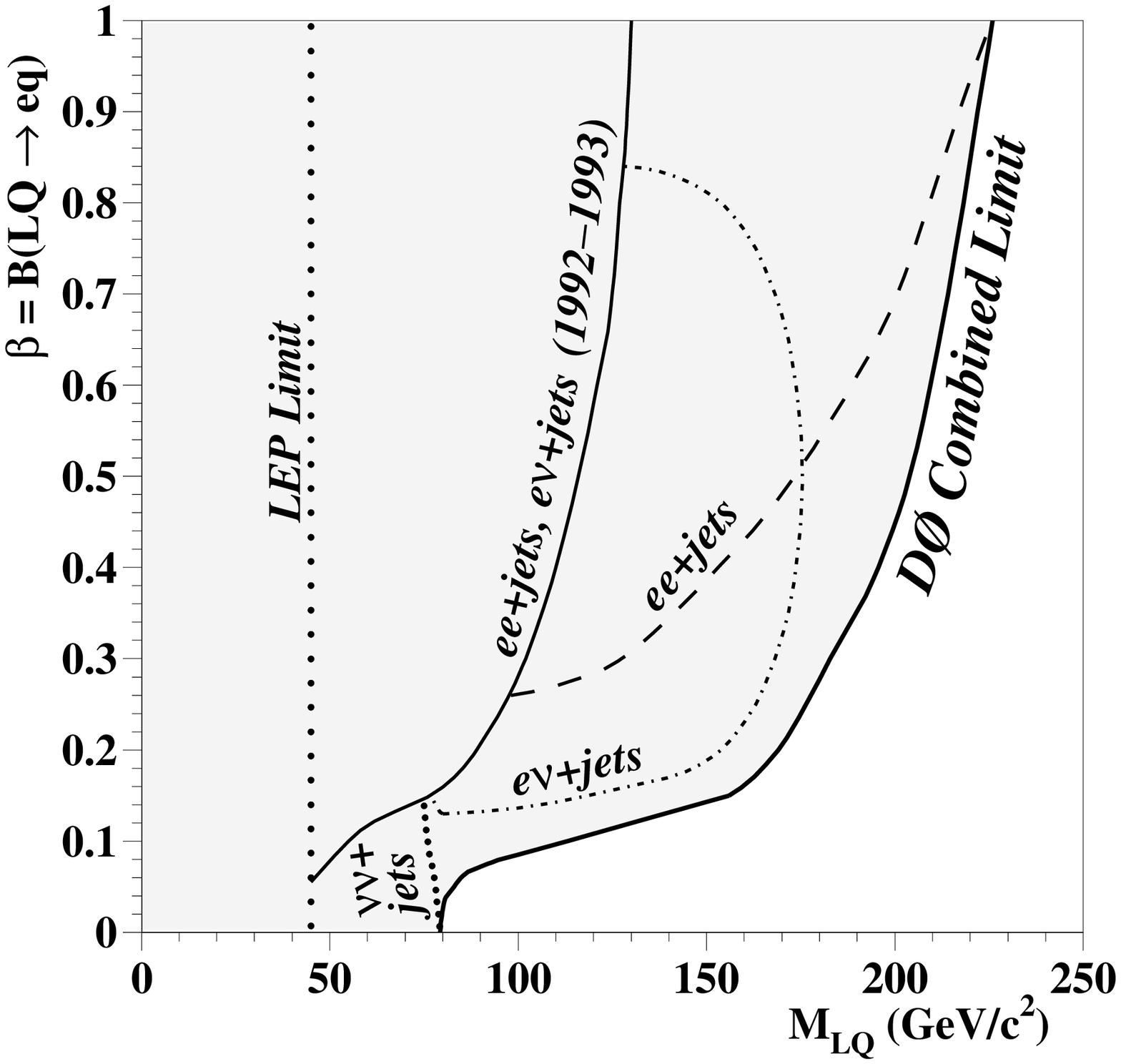,width=6.5in}}
\caption{ Lower limits on the first  generation scalar leptoquark mass as a
function of $\beta$, based on searches in all three possible decay channels
for LQ pairs. Limits from  LEP~\protect\cite{LEP} and
from our previous analysis~\protect\cite{old_lq2} of 1992--93 data are also
shown. The shaded area is excluded at 95\% CL. } \label{fig:exclusion}}
\end{figure}

Combining  the  $ee$+jets,   $e\nu$+jets, and   $\nu\nu$+jets  channels, we
calculate 95\% CL upper limits on the LQ pair production cross section as a
function of  LQ mass for  various  values of  $\beta$. These  cross section
limits for  $\beta =  \frac{1}{2}$  (shown in   Fig.~\ref{fig:limit}), when
compared with  NLO theory,  yield a 95\% CL  lower limit on  the LQ mass of
204~\gevcc.  The  lower  limits on  the LQ  mass  derived as a  function of
$\beta$, from all three  channels combined, as  well as from the individual
channels, are shown in  Fig.~\ref{fig:exclusion}. These results can also be
used to set limits on pair production of any heavy scalar particle decaying
into a lepton and a quark, in a variety of models.

In  conclusion, we  have  presented a  search for  first  generation scalar
leptoquark pairs  in the $e\nu +  {\rm jets}$ decay  channel. Combining the
results with  those from the  $ee + {\rm  jets}$ and  $\nu\nu + {\rm jets}$
channels, we exclude leptoquarks  with mass below 225~\gevcc\ for $\beta$ =
1, 204~\gevcc\ for  $\beta = {1 \over 2}$, and  79~\gevcc\ for $\beta = 0$,
at the 95\% CL. Our results exclude  (at the 95\% CL) the interpretation of
the HERA high $Q^2$ event excess via  $s$-channel scalar LQ production with
LQ mass below  200 \gevcc\  for values of  $\beta > 0.4$  and significantly
restrict new LQ models containing additional fermions~\cite{new-LQ}. 

We are grateful to  M.~Kr\"amer for discussions  and detailed cross section
information  and to J.L.~Hewett  and T.G.~Rizzo  for helpful discussions.
%
We thank the staffs at Fermilab and collaborating institutions for their
contributions to this work, and acknowledge support from the 
Department of Energy and National Science Foundation (U.S.A.),  
Commissariat  \` a L'Energie Atomique (France), 
State Committee for Science and Technology and Ministry for Atomic 
Energy (Russia),
CNPq (Brazil),
Departments of Atomic Energy and Science and Education (India),
Colciencias (Colombia),
CONACyT (Mexico),
Ministry of Education and KOSEF (Korea),
and CONICET and UBACyT (Argentina).

\end{document}